\documentstyle[epsf,twoside,fleqn,nuclphys]{article}
\def\photonatomright{\begin{picture}(3,1.5)(0,0)
                                \put(0,-0.75){\tencircw \symbol{2}}
                                \put(1.5,-0.75){\tencircw \symbol{1}}
                                \put(1.5,0.75){\tencircw \symbol{3}}
                                \put(3,0.75){\tencircw \symbol{0}}
                      \end{picture}
                     }

\def\photonatomup{\begin{picture}(1.5,3)(0,0)
                             \put(-0.75,3){\tencircw \symbol{3}}
                             \put(-0.75,1.5){\tencircw \symbol{2}}
                             \put(0.75,1.5){\tencircw \symbol{0}}
                             \put(0.75,0){\tencircw \symbol{1}}
                   \end{picture}
                  }

\def\photonright{\begin{picture}(30,1.5)(0,0)
                     \multiput(0,0)(3,0){10}{\photonatomright}
                  \end{picture}
                 }

\def\photonrighthalf{\begin{picture}(30,1.5)(0,0)
                     \multiput(0,0)(3,0){5}{\photonatomright}
                  \end{picture}
                 }

\def\photonuphalf{\begin{picture}(1.5,15)(0,0)
                      \multiput(0,0)(0,3){5}{\photonatomup}
                   \end{picture}
                  }

\def\fermionup{\begin{picture}(1,30)(0,0)
                     \put(0,0){\vector(0,1){15}}
                     \put(0,15){\line(0,1){15}}
               \end{picture}
              }
\def\fermionuphalf{\begin{picture}(1,15)(0,0)
                         \put(0,0){\vector(0,1){7.5}}
                         \put(0,7.5){\line(0,1){7.5}}
                   \end{picture}
                  }

\def\fermionull{\begin{picture}(30,15)(0,0)
                        \put(0,0){\vector(-2,1){15}}
                        \put(-15,7.5){\line(-2,1){15}}
                  \end{picture}
                 }
\def\fermionullhalf{\begin{picture}(15,7.5)(0,0)
                        \put(0,0){\vector(-2,1){7.5}}
                        \put(-7.5,3.75){\line(-2,1){7.5}}
                  \end{picture}
                 }
\def\fermionurr{\begin{picture}(30,15)(0,0)
                        \put(-30,-15){\vector(2,1){15}}
                        \put(-15,-7.5){\line(2,1){15}}
                  \end{picture}
                 }
\def\fermionurrhalf{\begin{picture}(15,7.5)(0,0)
                        \put(-15,-7.5){\vector(2,1){7.5}}
                        \put(-7.5,-3.75){\line(2,1){7.5}}
                  \end{picture}
                 }

\newenvironment{Feynman}[3]{\begin{center}
                            \setlength{\unitlength}{#3 mm}
                            \begin{picture}(#1)(#2)
                            \thicklines
                           }{\end{picture} \end{center}}

\newcommand{\nll}{\nonumber \\}

\newcommand{\bq}{\begin{equation}}
\newcommand{\eq}{\end{equation}}
\newcommand{\ba}{\begin{eqnarray}}
\newcommand{\ea}{\end{eqnarray}}

\newcommand{\mr}{\mathrm}
\newcommand{\mathrm}{\rm}
\newcommand{\ltwo}{LEP~2}

\begin{document}
\thispagestyle{empty}
\title{
\, \vspace{-2.5cm}
\large
DESY 94--
\hfill    {\tt ISSN 0418-9833}
\\
CERN--TH. 7305/94
\\
LMU 08/94
\\
June 1994
\vspace{2.cm}
\\
\LARGE \bf
Semi-analytical approach to Higgs production at LEP~2\thanks
{
Talk presented by A. Leike
at the Zeuthen Workshop on Elementary
Particle Theory -- Physics at LEP200 and Beyond, held at Teupitz,
Germany, 10--15 April 1994; to appear in the proceedings.
}
\vspace*{0.5cm}
}
\author{
Dima Bardin\address{Joint Institute for Nuclear Research, ul. Joliot-Curie 6,
RU--141980 Dubna, Moscow Region, Russia
}\address{Theoretical Physics Division, CERN, CH--1211 Geneva 23, Switzerland
}\thanks{Supported by the Ministry for Science, Research, and Culture of
Land Brandenburg, contract II.1-3141-2/8(94).},
Arnd~Leike\address{Lehrstuhl Prof. Fritzsch, Sektion Physik der
Ludwig-Maximilians-Universit\"at,
Theresienstr. 37, D--80333 M\"unchen, Germany}\thanks{
Supported by the German Federal Ministry for Research
and Technology under contract No.~05~GMU93P.}
        and
Tord Riemann\address{DESY--Institut f\"ur Hochenergiephysik, Platanenallee 6,
D--15738
Zeuthen, Germany}
}
\begin{abstract}
The cross-section for the reaction $e^+e^- \rightarrow b\bar{b}\mu^+\mu^-$ is
calculated
with a
semi-analytical integration of the phase space.
Compact formulae are obtained
for the total cross
section and for
invariant mass distributions of the $\mu^+\mu^-$ and $b\bar{b}$ pairs.
The background diagrams to $ZH$ production yield analytically
 cumbersome but numerically small contributions.
The numerical results are compared with those
from a Monte Carlo approach.
\end{abstract}
\maketitle

%
\section{INTRODUCTION}
A main task of LEP~2
 will be the investigation of
the production of two
heavy gauge bosons.
As it is well
known, these particles are extremely unstable.
Immediately after their production, they decay preferably into
a four-fermion final state.
The same final states are produced by
 competing background reactions with one or no gauge boson in the
intermediate states.

A spectacular event could be the observation of Higgs production at LEP~2
via the Bjorken process~\cite{bjorken}:
\ba
\label{bj}
e^+e^- \rightarrow Z H. 
\ea
For the Standard Model Higgs mass range of interest at LEP, it is $M_H < 2M_W$
and
the Higgs boson decays with a probability of
almost 100\% into a pair of $b$-quarks.
Therefore, the competing reactions to Higgs
production at LEP~2 are four-fermion final states containing a $b$-quark
pair.
Monte Carlo calculations of the processes
\ba
e^+e^- &\rightarrow& Z b \bar{b},
\\
\label{bm}
e^+e^- &\rightarrow& \mu^+ \mu^- b \bar{b}
\ea
have been performed in~\cite{boos}.

\bigskip

In this contribution, we apply our semi-analytical approach~\cite{bardin}
to the calculation of  the cross-section of reaction~(\ref{bm}).
However, our analytical results are also applicable to
a full class of similar processes:
\ba
\label{4f}
e^+e^- \rightarrow f_1\bar{f_1}f_2\bar{f_2},
\ea
with $f_i\ne e,\nu$ and $f_1\ne f_2$.
%
\section{AMPLITUDES \mbox{AND INTERFERENCES}}

\begin{figure*}
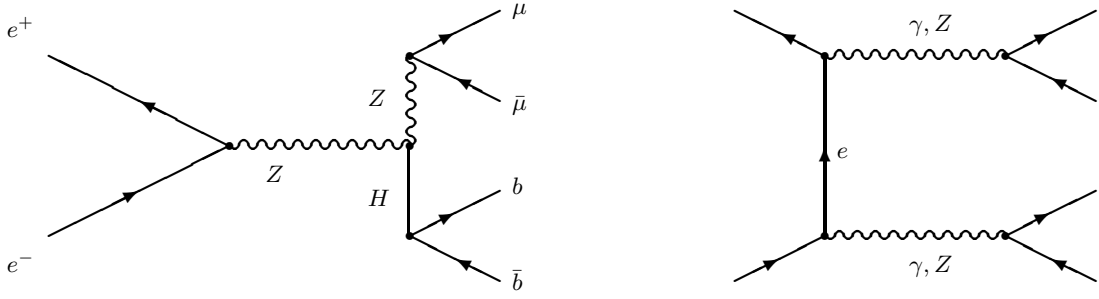

\begin{minipage}[tbh]{7.8cm}{
\begin{center}
\begin{Feynman}{75,60}{0,0}{0.8}
%
\put(30,30){\fermionurr}
\put(30,30){\fermionull}
\put(30,30){\photonright}
\put(30,30){\circle*{1.5}}
\put(60,30){\circle*{1.5}}
\put(60,15){\circle*{1.5}}
\put(60,45){\circle*{1.5}}
\put(75,7.5){\fermionullhalf}
\put(75,22.5){\fermionurrhalf}
\thinlines
\put(59.75,15){\line(0,1){15}}
\thicklines
\put(60,30){\photonuphalf}
\put(75,37.5){\fermionullhalf}
\put(75,52.5){\fermionurrhalf}
\small
\put(-07,48){$e^+$}  
\put(-07,09){$e^-$}
\put(36,24){$Z$}
\put(53,36.5){$Z$}  
\put(53,20.0){$H$}
\put(77,06){${\bar b}$}  
\put(77,22){$ b$}
\put(77,36){${\bar \mu}$}  
\put(77,52){$\mu$}
\normalsize
\end{Feynman}
\end{center}
}\end{minipage}
\begin{minipage}[tbh]{7.8cm} {
\begin{center}
\begin{Feynman}{75,60}{0,0}{0.8}
%
\put(30,15){\fermionurrhalf}
\put(30,45){\fermionullhalf}
\put(30,15){\fermionup}
\put(30,45){\photonright}
\put(30,15){\photonright}
\put(30,45){\circle*{1.5}}
\put(30,15){\circle*{1.5}}
\put(60,15){\circle*{1.5}}
\put(60,45){\circle*{1.5}}
\put(75,7.5){\fermionullhalf}
\put(75,22.5){\fermionurrhalf}
\put(75,37.5){\fermionullhalf}
\put(75,52.5){\fermionurrhalf}
\small
\put(32,28){$e$}
\put(44,49){$\gamma,Z$}  
\put(44,09){$\gamma,Z$}
\normalsize
\end{Feynman}
\end{center}
}\end{minipage}
\caption
{
\label{higgs}
The basic diagram for off-shell $ZH$ production~(a)
and the {\tt crab} type background~(b).
}
\end{figure*}

The process~(\ref{bm})
is described by seven generic diagrams:
\begin{itemize}
\item[(i)]
The Higgs signal diagram, figure~1a;
\item[(ii)]
$t$ and $u$ channel exchange background diagrams of the {\tt crab} type,
figure~1b;
\item[(iii)]
four non-resonating background diagrams of the {\tt (rein)deer} type, figure~2.
\end{itemize}
Every of the generic diagrams in (ii) and (iii)
represents four Feynman diagrams with the
virtual neutral gauge bosons
being photons or $Z$ bosons.
Thus, the whole process is described by $1+4\cdot 2+4\cdot 4=25$
Feynman diagrams.
The complexity of the problem arises from the properties of the seven
types of generic diagrams.
Here we only mention that the interferences of the Higgs diagram
with all the others vanish after an integration over the angles
characteristic of the $b$-quark pair -- with one exception:
the two {\tt deer} diagrams with the $b$-quarks being attached to the
first intermediate $\gamma,Z$ in figure~2 ($f_1= b$).
All the other interferences with the signal
yield the same trace over the $b$-quark line:
\ba
Tr\left[ (\rlap/p_b+m)(\rlap/p_{\bar{b}}-m)\gamma^\alpha\right]
= 4m(p_{\bar{b}}-p_b)^\alpha.
\ea
After an integration over
the phase space of the $b$-quarks:
\ba
\sigma^{\mr{int}} &=& \int \mbox{d}\Omega_b (p_{\bar{b}}-p_b)^\alpha
T_\alpha(k_1,k_2,p_{\mu^+},p_{\mu^-})
\nll
&=& 0.
\ea
Here it is essential that $T_\alpha$ is independent of the momenta
$p_{\bar{b}}$ and
$p_b$.
So, the integrand changes sign when interchanging $p_{\bar{b}}$
and $p_b$ but $d\Omega_b$ does not.
As a result, the integral vanishes
after the integration over the $b$-quark angles.

Thus, the Higgs signal adds up incoherently with the
{\tt crab} contributions and a large fraction
of the other background
if not $b$-quark asymmetries are studied.

In addition, we would like to mention that
all the interferences of the Higgs signal with
background diagrams are suppressed with respect to the squared Higgs
diagram
due to the narrow width of the Higgs boson.
A rough estimate may be obtained as follows.
Be $\chi_B(s)$ a boson propagator,
\ba
\label{chi}
\chi_B(s)=\frac{s}{s-M_B^2+i\Gamma_BM_B}.
\ea

If there is a chance at all to find
a Higgs boson at \ltwo,
then it is not unrealistic to
assume both Higgs boson and the $Z$ to be nearly on their mass shells.
Then, the ratio of the propagators may be estimated to be roughly as
follows:
\ba
\label{frac}
\frac{\chi_Z}{\chi_H} \approx \frac{\Gamma_Z}{\Gamma_H},
\hspace{.7cm}
\frac{\chi_\gamma}{\chi_H} \approx \frac{M_H}{\Gamma_H}.
\ea

\begin{figure*}
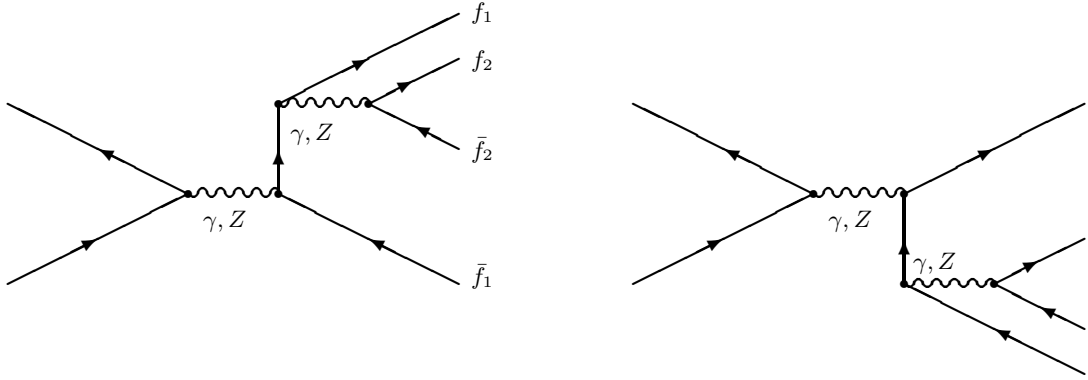

\begin{minipage}[tbh]{7.8cm}{
\begin{center}
\begin{Feynman}{75,60}{-5.0,0}{0.8}
%
\put(20,30){\fermionurr}
\put(20,30){\fermionull}
\put(20,30){\photonrighthalf}
\put(65,15){\fermionull}
\put(20,30){\circle*{1.5}}
\put(35,30){\circle*{1.5}}
\put(35,45){\circle*{1.5}}
\put(50,45){\circle*{1.5}}
\put(35,30){\fermionuphalf}
\put(65,60){\fermionurr}
\put(35,45){\photonrighthalf}
\put(65,52.5){\fermionurrhalf}
\put(65,37.5){\fermionullhalf}
\small
\put(22.5,24.5){$\gamma, Z$}
 \put(37,39){$\gamma,Z$}
\put(67,59){$f_1$}
\put(67,51){$f_2$}
\put(67,36){${\bar f}_2$}  
\put(67,15){${\bar f}_1$}  
\normalsize
\end{Feynman}
\end{center}
}\end{minipage}
\begin{minipage}[tbh]{7.8cm}{
\begin{center}
\begin{Feynman}{75,60}{0,0}{0.8}
%
\put(30,30){\fermionurr}
\put(30,30){\fermionull}
\put(30,30){\photonrighthalf}
\put(30,30){\circle*{1.5}}
\put(45,30){\circle*{1.5}}
\put(45,15){\circle*{1.5}}
\put(60,15){\circle*{1.5}}
\put(45,15){\fermionuphalf}
\put(75,00){\fermionull}
\put(75,45){\fermionurr}
\put(45,15){\photonrighthalf}
\put(75,22.5){\fermionurrhalf}
\put(75,07.5){\fermionullhalf}
\small
\put(32.5,24.5){$\gamma, Z$}
 \put(46.5,17.5){$\gamma,Z$}
\normalsize
\end{Feynman}
\end{center}
}\end{minipage}
\caption{
\label{ders}
Background contributions to off-shell $ZH$ production:
up and down {\tt reindeers}; $f_i=\mu, b$.
}
\end{figure*}

Below the threshold of the decay $H \rightarrow W^+W^-$,
the off-shell width of the $H$ boson is
\bq
\label{ghoff}
\Gamma_H (s) =
  \frac{G_{\mu}} {4\pi \sqrt{2}} \sqrt{s} \sum_f m_f^2 N_c(f),
\eq
and that of the $Z$ is
\bq
\label{gzoff}
\Gamma_Z (s) =
  \frac{G_{\mu}\, M_Z^2} {24\pi \sqrt{2}} \sqrt{s}
  \sum_f (v_f^2+a_f^2)N_c(f),
\eq
where $N_c(f)=1\ (3)$ for leptons (quarks).
We use the normalization $a_f=1$.

For $M_H < 2 M_W$
the Higgs width is of the order of
a few MeV and
the non-vanishing interferences of background with the Higgs signal
are highly suppressed at LEP~2.
%
\section{\mbox{PHASE SPACE}
\mbox{AND CROSS SECTIONS}}
We parametrize the eight-dimensional  phase space of four final state
particles as follows:
\begin{eqnarray}
&&d\Omega
= \prod_{i=1}^4\frac{d^3p_i}{2p_{i}^0}
\delta^4(k_1+k_2-\sum_{i=1}^4 p_i)
\\
&=&~2\pi\frac{\sqrt{\lambda(s,s_H,s_Z)}}{8s}
\frac{\sqrt{\lambda(s_H,m_b^2,m_b^2)}}{8s_H}
\nll
&\times&~\frac{\sqrt{\lambda(s_Z,m_{\mu}^2,m_{\mu}^2)}}{8s_Z}
d s_H d s_Z d \cos\theta d \Omega_H d \Omega_Z,
\nonumber
\end{eqnarray}
where we already integrated over the rotation angle around the beam axis.
The $k_1$ and $k_2$ are the four-momenta of the initial
electron and positron and $p_i$ those of the final state particles.
The invariants $s, s_H$, and $s_Z$ are
\begin{eqnarray}
s&=&(k_1+k_2)^2,
\nll
s_H&=&(p_1+p_2)^2,
\hspace{0.5cm}
 s_Z=(p_3+p_4)^2,
\end{eqnarray}
and $\theta$ is the angle between the vectors ($\vec{p}_1+\vec{p}_2$) and
$\vec{k}_1$.
The spherical angles
of the momenta $\vec{p_1}$ and $\vec{p_2}$ ($\vec{p_3}$ and $\vec{p_4}$)
in their rest frames are in
$\Omega_H\ (\Omega_Z)$:
$d \Omega_i = d \cos\theta_i d \phi_i$.
The kinematical ranges of the integration variables are:
\begin{eqnarray}
(2m_b)^2  &\le& s_H    \le (\sqrt{s}-2m_{\mu})^2,
\nll
(2m_{\mu})^2 &\le& s_Z \le (\sqrt{s}-\sqrt{s_H})^2,
\nll
-1 &\le& \cos\theta,\ \cos\theta_H,\ \cos\theta_Z \le 1,
\nll
0 &\le& \phi_H,\phi_Z \le 2\pi.
\end{eqnarray}
We integrated analytically over all the angular variables, leaving
the integrations over $s_H$ and $s_Z$ to be performed numerically.

The cross section is:
\begin{eqnarray}
\label{sig}
\sigma(s)=\int_{{\bar s}_H}^s d s_H \rho(s_H)
\int_{{\bar s}_Z}^{(\sqrt{s}-\sqrt{s_H})^2} d s_Z\\ \nonumber
\times \rho(s_Z) \sigma_0(s,s_H,s_Z),
\end{eqnarray}
with
\begin{eqnarray}
\rho(s)=\frac{1}{\pi}\frac{\sqrt{s} \, \Gamma(s)}
{|s-M^2+i\sqrt{s} \, \Gamma(s)|^2}\cdot {\mr{BR}}.
\end{eqnarray}
The
$M$ and $\Gamma$ are mass and width of the resonating off shell particles
and ${\mr{BR}}$ is the branching ratio of its decay to the observed final state
fermion pair.
The
$\rho(s)$ has the property
$\lim_{\Gamma\rightarrow 0}\rho(s)\rightarrow \delta(s-M^2){\mr{BR}}$.
The lower integration bounds ${\bar s}_H$ and ${\bar s}_Z$
cut on the invariant masses of the $b$-quark and muon pairs.

The functions $\sigma_0(s,s_H,s_Z)$ in~(\ref{sig}) are the result of a
fivefold analytical integration.
The result has the following structure:
\begin{eqnarray}
\label{sig1}
\sigma_{0}(s,s_H,s_Z)
&=&
\sigma_{0}^{H} + \sigma_{0}^{H,b-{\tt{deers}}}
\nll
+~\sigma_{0}^{{\tt{crab}}}
&+& \sigma_{0}^{{\tt{deers}}}
+ \sigma_{0}^{{\tt{crab}},{\tt{deers}}}.
\end{eqnarray}
Here, the superscripts denote the different interferences of the generic
diagrams.
The numerically largest contributions are
\ba
\label{sigh}
\sigma_0^H(s;s_1,s_2)  =
\frac{\left(G_{\mu} M_Z^2 \right)^2} {96 \pi s}
\frac{M_Z^2}{s} \left( v_e^2+a_e^2 \right)
\nll
\times \left| \frac{s}{s-M_Z^2+i M_Z\Gamma_Z(s)} \right|^2
{\cal G}_4^{\mr{Bj}}(s;s_1,s_2)
\ea
with the kinematical function
\bq
\label{zhmuta}
{\cal G}_4^{\mr {Bj}}(s;s_1,s_2) =
\frac{\lambda^{1/2}}{s^2 s_2}
\left(\lambda +12 s s_2\right)
\eq
and~\cite{bardin}
\ba
\label{sigz}
\lefteqn
{
\sigma_{0}^{{\tt{crab}}}(s;s_1,s_2) =
\biggl[
 \frac{\left(G_{\mu} M_Z^2 \right)^2}{64\pi s} \!
\left(v_e^4+6v_e^2a_e^2+a_e^4\right)
}
\nll
\hspace*{-.5cm}&
+~ \ldots \biggr]
{\cal G}_4^{\mr{t+u}}(s,s_1,s_2)
\ea
with
\ba
{\cal G}_4^{\mr{t+u}}(s;s_1,s_2) =
\frac{\lambda^{1/2}}{s}\left[\frac{s^2+(s_1+s_2)^2}{s-s_1-s_2}
  {\cal L}_4 -2 \right].
\nonumber 
\ea
The dots in~(\ref{sigz}) indicate the contributions with intermediate
photons. Further,
the following definitions are used:
\ba
\label{lambda}
\lambda &\equiv&
 \lambda(s;s_1,s_2)
\nll
&=&
s^2 + s_1^2 + s_2^2 - 2ss_1 - 2 s_1s_2 - 2 s_2s
\ea
and
\ba
\label{L4}
{\cal L}_4(s;s_1,s_2) &=&
\frac{1}{\sqrt{\lambda}} \, \ln \frac{s-s_1-s_2+\sqrt{\lambda}}
                                     {s-s_1-s_2-\sqrt{\lambda}}~.
\ea
%
The results for the {\tt deer} diagrams and their interferences are rather
lengthy and will be published elsewhere\footnote
{
The {\tt crab} and {\tt deer} cross sections alone have been calculated
in a Monte Carlo approach in~\cite{pittau}.
}.

The above results may be used for the
calculation not only of
total cross sections but
also of distributions
$\mbox{d}\sigma/\mbox{d}s_H,\  \mbox{d}\sigma/\mbox{d}s_Z$, and
$\mbox{d}\sigma/\mbox{d}\cos\theta$ which are of vital
importance for a Higgs search~\cite{felcini}.
Further,
$b$-quark and muon pairs of low invariant mass may be cut in order
to avoid nonperturbative effects due to bound states and
large background due to photon exchange.

The equations~(\ref{sig}) and~(\ref{sigh})--(\ref{sigz}) contain the
numerically
important contributions to the reaction
$e^+e^-\rightarrow b\bar{b}\mu^+\mu^-$ at the Born level.
For applications to
data, the inclusion of the universal initial state
QED corrections is important~\cite{convol}:
\ba
\sigma_T = \int_{\bar s}^s \, \frac{d s'}{s} \, \sigma(s') \, \rho(s'/s),
\ea
with ${\bar s} \geq (2m_{\mu}+2m_b)^2$.
The accuracy then should be quite sufficient for a search experiment
although certain QED and weak corrections are left
out~\cite{bardin,bbor,kniehl}.
%
\section{RESULTS}
The total cross section $\sigma(s)$
is
shown in figure~3 for various Higgs masses.
Cuts on the invariant masses have been applied:
$\sqrt{s_H}=E_{b{\bar b}}\geq 12$ GeV, $\sqrt{s_Z}=E_{\mu^+\mu^-}
\geq 12$ GeV.
We found agreement with results from a Monte Carlo
calculation~\cite{boos} within the errors of the latter.
With an integrated luminosity of $500\,fb^-1$ one may expect
 between 10 and 20
$b\bar{b}\mu^+\mu^-$ events.

The $b$-quark pairs from Higgs decay
have a $\delta$-function like peak in
the energy distribution, see figure~4.
This is due to the small Higgs width which may be estimated in the
Standard Model to about
4, 5, 6\, MeV for $M_H$ = 80, 100 and 140\,GeV.
This way, $b$-quark pairs from Higgs decay may easily be
distinguished from background even if $M_H\approx M_Z$.
Although a Higgs with a mass of 120\,GeV could give a strong peak in the
energy distribution of the
$b$-pairs, it would be too heavy to be detected at \ltwo\
because it gives less then 1 event for the considered luminosity even if
$\sqrt{s}=200$ GeV would be realized; see figure~5.
The maximal mass for which a Higgs boson detection seems feasible at
\ltwo\
is $M_H\approx \sqrt{s}-100\,GeV$~\cite{felcini}.

\bigskip

\begin{figure}[htbp]
\begin{center}
\vspace{-0.5cm}
\hspace{-2.0cm}
\mbox{
\epsfysize=7.0cm
\epsffile[0 0 500 500]{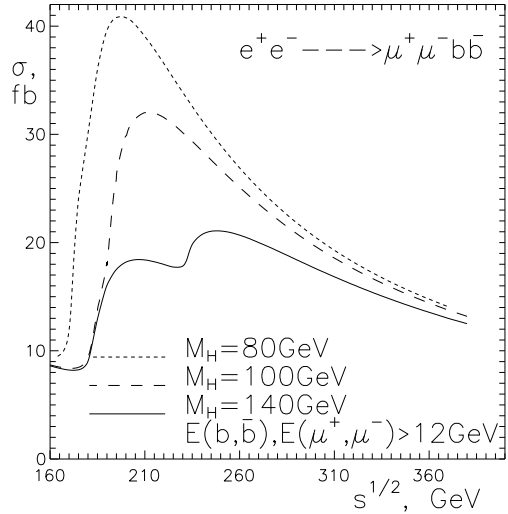}}
\end{center}
\vspace{-1.2cm}
\caption[]
{
\label{f3}
The total cross section $\sigma(e^+e^- \rightarrow b\bar{b}\mu^+\mu^-)$
as function of $s^{1/2}$ for various Higgs masses.
}
\end{figure}

\vfill

\begin{figure}[bhtp]
\vspace{-0.5cm}
\begin{center}
\hspace{-1.4cm}
\mbox
{
\epsfysize=7cm
\epsffile[0 0 500 500]{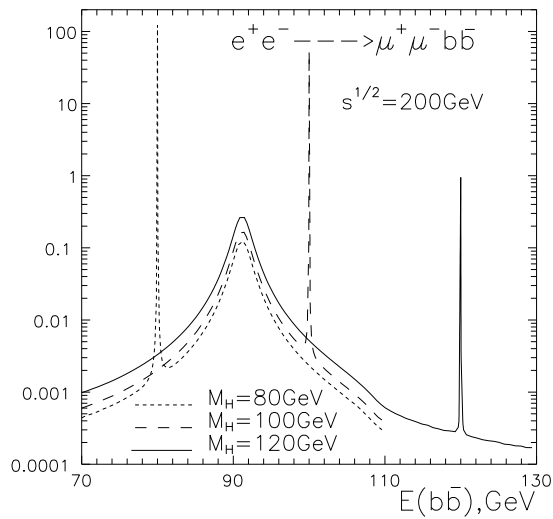}
}
\end{center}
\vspace{-1.2cm}
\caption{
\label{f4}
The distribution
$\mbox{d}\sigma/\mbox{d}E_{b{\bar b}}\cdot E_{beam}/\sigma$
for different Higgs masses at $s^{1/2}=200$\,GeV.
A cut $E_{\mu^+\mu^-}\geq 12$ GeV has been applied.
}
\end{figure}

\begin{figure}[thbp]
\begin{center}
\vspace{-0.5cm}
\hspace{-2.3cm}
\mbox{
\epsfysize=7cm
\epsffile[0 0 500 500]{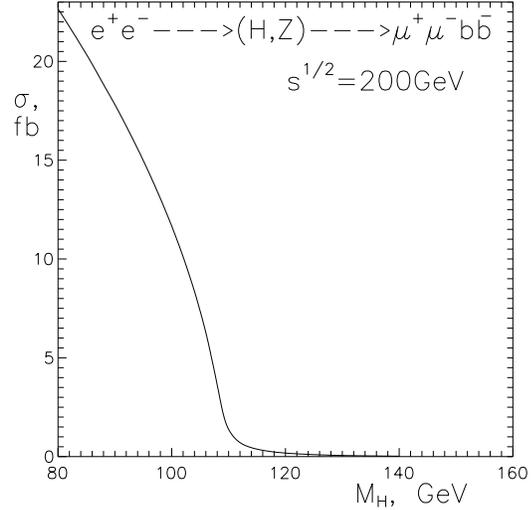}
}
\end{center}
 \vspace{-1.2cm}
\caption{
\label{f5}
The total cross section
$\sigma(e^+e^- \rightarrow (H,Z)\rightarrow b \bar{b}\mu^+\mu^-)$
 as a function of the Higgs mass for $s^{1/2}=200$\,GeV.
}
\end{figure}

 {\em To summarize},
we performed the first complete semi-analytical calculation
of the off-shell Bjorken process, $e^+e^-\rightarrow b\bar{b}\mu^+\mu^-$.
 The interferences
between the Higgs signal and the resonating {\tt crab}
background are zero after integration
over the angles of the $b$-quark pair.
The rest of the background may be neglected in a search experiment.
The analytical results are applicable also for reactions of the type
$e^+e^-\rightarrow f_1\bar{f}_1f_2\bar{f}_2$, where $f_1\neq f_2$ and
$f_1,f_2\neq e,\nu_e$.
This opens the possibility for a further study of e.g. the higher order
fermion pair corrections to the $Z$ line shape~\cite{kkks}.
\section*{ACKNOWLEDGEMENT}
We would like to thank E. Boos for a discussion.


\end{document}